\documentclass[12pt,twoside]{article}   %For printing on two sides of page
\usepackage[super,sort,comma]{natbib}
\usepackage{amsfonts}
\usepackage{amsmath}
\usepackage{epsfig}
\usepackage{graphicx}
\usepackage{comment}
\usepackage{amsmath,amssymb} % define this before the line numbering.
\usepackage{color}
\usepackage{subfigure}
\usepackage{animate}
\usepackage{float}
\usepackage{bm}
\usepackage{multirow}
\usepackage{rotating}
\usepackage{wrapfig}
\usepackage{epsfig}
\usepackage{nicefrac}       % compact symbols for 1/2, etc.
\usepackage{microtype} 
\usepackage{float}
\usepackage{amsmath}
\usepackage{enumerate}
\usepackage{enumitem} 
\usepackage{amsmath,amsfonts,amssymb}
\usepackage{graphicx}
\usepackage[colorlinks=true, allcolors=blue]{hyperref}
\usepackage{amsfonts}
\usepackage{graphicx}
\usepackage{amsmath,amssymb}
 \usepackage{xcolor} 
\usepackage{appendix}
\usepackage{fancyhdr}		%Gives headers and footers defined below
\usepackage[section]{placeins}   %
\usepackage{graphicx}

\makeatletter \renewcommand\@biblabel[1]{$^{#1}$} \makeatother
\newcommand{\cen}[1]{\begin{center} #1 \end{center}}

\usepackage[mathlines]{lineno}
\usepackage{hyperref}
\hypersetup{ colorlinks,
    citecolor=blue,
    filecolor=blue,
    linkcolor=blue,
    urlcolor=blue
}

\usepackage{xcolor}
\definecolor{gray}{rgb}{0.6,0.6,0.6}
\definecolor{red}{rgb}{0.85,0,0}
\definecolor{green}{rgb}{0,0.85,0}
\definecolor{blue}{rgb}{0,0,0.85}
\definecolor{beige}{rgb}{0.92,0.87,0.78}
\usepackage[all]{hypcap}    %causes link to figures to go to figure, not caption
\begin{document}

\cen{\sf {\Large {\bfseries {Posterior Estimation Using Deep Learning: A Simulation Study of Compartmental Modeling in Dynamic PET}} \\  
\vspace*{10mm}
Xiaofeng Liu, Thibault Marin, Tiss Amal, Jonghye Woo, Georges El Fakhri, Jinsong Ouyang} \\~
\\
Gordon Center for Medical Imaging, Radiology Department, Massachusetts General Hospital, Boston, MA 02114, USA.

Radiology Department, Harvard Medical School, Boston, MA 02115, USA.

\vspace{5mm}
%Version typeset \today\\
}

\pagenumbering{roman}
\setcounter{page}{1}
\pagestyle{plain}
Corresponding to: ouyang.jinsong@mgh.harvard.edu \\
% note, probably best not to use a student's e-mail as it won't be valid for
% very long.

\begin{abstract}
\noindent {\bf Background:} In medical imaging, images are usually treated as deterministic, while their uncertainties are largely underexplored. \\ 
{\bf Purpose:} This work aims at using deep learning to efficiently estimate posterior distributions of imaging parameters, which in turn can be used to derive the most probable parameters as well as their uncertainties.\\
{\bf Methods:} Our deep learning-based approaches are based on a variational Bayesian inference framework, which is implemented using two different deep neural networks based on conditional variational auto-encoder (CVAE), CVAE-dual-encoder and CVAE-dual-decoder. The conventional CVAE framework, i.e., CVAE-vanilla, can be regarded as a simplified case of these two neural networks. We applied these approaches to a simulation study of dynamic brain PET imaging using a reference region-based kinetic model.\\
{\bf Results:} In the simulation study, we estimated posterior distributions of PET kinetic parameters given a measurement of time-activity curve. Our proposed  CVAE-dual-encoder and CVAE-dual-decoder yield results that are in good agreement with the asymptotically unbiased posterior distributions sampled by Markov Chain Monte Carlo (MCMC). The CVAE-vanilla can also be used for estimating posterior distributions, although it has an inferior performance to both CVAE-dual-encoder and CVAE-dual-decoder.\\
{\bf Conclusions:} We have evaluated the performance of our deep learning approaches for estimating posterior distributions in dynamic brain PET. Our deep learning approaches yield posterior distributions, which are in good agreement with unbiased distributions estimated by MCMC. All these neural networks have different characteristics and can be chosen by the user for specific applications. The proposed methods are general and can be adapted to other problems. \\

%\noindent This work is to use deep learning for efficiently estimating posterior distributions of imaging parameters, which in turn can be used to derive the most probable parameters as well as their uncertainties. Our deep learning-based approaches are based on a variational Bayesian inference framework. We applied our approaches to a simulation study of dynamic brain PET imaging using a reference region-based kinetic model. In the simulation study, we estimated posterior distributions of PET kinetic parameters. Our approaches yield results that are in good agreement with the asymptotically unbiased posterior distributions sampled by Markov Chain Monte Carlo.\\

\end{abstract}

\newpage     %may or may not be needed

%The table of contents is for drafting and refereeing purposes only. Note
%that all links to references, tables and figures can be clicked on and
%returned to calling point using cmd[ on a Mac using Preview or some
%equivalent on PCs (see View - go to on whatever reader).
%\tableofcontents

\newpage

\setlength{\baselineskip}{0.7cm}      %double spacing		

\pagenumbering{arabic}
\setcounter{page}{1}
\pagestyle{fancy}

\section{Introduction}
\label{sec:introduction}

Uncertainty quantification of medical imaging data is fundamentally important for clinical diagnosis and clinical trials. However, medical images presented in both research and clinical settings are usually non-statistical in the sense that they do not contain information about uncertainty. From the point of view of statistical inference, this belongs to the frequentist method \cite{cox2006principles}, in which images are treated as deterministic. To assess the uncertainty, frequentist inference requires repeated measurements, which is impractical for medical imaging. Without uncertainty information, assessment of research results and clinical images can be challenging and, under certain circumstances, lead to incorrect conclusions and clinical decisions. However, we usually have prior knowledge of the image to be estimated before the measurement is made. Such prior knowledge can be combined with the measurement to obtain an estimation of the posterior distribution, which can then be used to assess the uncertainty of the image. This falls into Bayesian inference (BI) \cite{cox2006principles}, which is a coherent solution to the problem of uncertainty estimation. {The posterior distribution is a full distribution on the parameter. It is possible to make all sorts of probabilistic statements about the parameter. For example, we can make a statement of credible interval (in contrast to confidence interval in the frequentist method) if a posterior distribution is known.}      
 
Most medical imaging problems can be generalized as the estimation of $\bm{x}$ in a parameter space given an observable measurement $\bm{y}$. In the framework of BI, we define the problem as: given $\bm{y}$ and a prior, $p(\bm{x})$, which represents our knowledge on $\bm{x}$ before the measurement, what is the posterior distribution, $p(\bm{x}|\bm{y})$? The conventional method to tackle this problem is to use Markov Chain Monte Carlo (MCMC) \cite{andrieu2003introduction}, which is known to produce an asymptotically unbiased estimation of the posterior distribution. MCMC does not require a full analytic posterior description as long as the ratios of probability density functions at pairs of locations (i.e., $\bm{x}$’s) can be calculated \cite{hastings1970monte}. Although this requirement is met for many medical imaging problems, MCMC has been rarely used in the past. One reason is that recomputing likelihood $p(\bm{y}|\bm{x})$ becomes too expensive for most problems without even accounting for the fact that a large number of burn-in steps are needed for MCMC. {Using a dynamic positron emission tomography (PET) study performed on a GE Discovery MI-5 scanner as an example, a time series of sinogram set, $\bm{y}$, has a dimension of 54×1981×415×272 (assuming 54 time frames; the scanner uses 1981 sinograms in each sinogram set, while each sinogram has 415 radial bins and 272 angular bins), while the corresponding time series of image volumes, $\bm{x}$, has a dimension of 54×256×256×256 (No time of flight is considered here. Otherwise, another dimension of 31 will be added to $\bm{y}$). Approximate Bayesian Computing (ABC) is another well-known method for estimating posterior distributions for a given measurement \cite{fan2021pet}.  In ABC, model parameters sampled from the prior are used to generate artificial measurement datasets. If the resulting datasets are very close to the given measurement according to a predefined discrepancy function, the corresponding parameters are then accepted as the part of the posterior. Unlike MCMC, ABC is an approximation. Also, it does not offer much advantage over MCMC in terms of computational time.} Another reason is that the prior is subjective in BI. A poor prior certainly leads to a poor posterior estimation.

As a large amount of training data becomes available in medical imaging, BI combined with deep learning (DL) has the potential to play an important role in posterior estimation in the future. We first define a very general problem, which is not limited to medical imaging, as:

%, where $i=1,\cdots,D$
 
\textit{Given a training dataset of $D$ samples, $\{\bm{x}_i,\bm{y}_i\}_{i=1}^D$, which represents a forward mapping from parameter $\bm{x}$ to measurement $\bm{y}$, and a testing observable measurement, $\bm{y}^*$, what is the posterior distribution, $p(\bm{x}|\bm{y}^*)$?}

\noindent In this definition, the prior, $p(\bm{x})$, is no longer subjective but implicitly defined by the training dataset itself. To solve the above problem using MCMC is challenging because only training data rather than the underlying analytic forward and noise models are available. This makes it difficult to compute the ratios of probability density functions at pairs of locations as required by MCMC. {For such a problem, it is also difficult to use ABC from the training data without knowing the underlying model.} We intend to sample the posterior distribution, $p(\bm{x}|\bm{y}^*)$, using a conditional variational auto-encoder (CVAE), in which the generation process is conditioned on $\bm{y}^*$. In addition, we introduce a latent multivariate random variable $\bm{z}$ to account for the information loss in the forward process from $\bm{x}$ to $\bm{y}$ \cite{ardizzone2019analyzing}. The CVAE is trained with the paired dataset $\{\bm{x}_i,\bm{y}_i\}_{i=1}^D$. The trained decoder in the CVAE can then be used to generate the posterior distribution, $p(\bm{x}|\bm{y}^*)$, which represents a complete picture of the parameter space, using a predefined distribution of the latent variable, $p(\bm{z})$. Based on this strategy, we have derived different DL-based approaches for estimating posterior distributions using the CVAE framework (See Sec. \ref{sec:method}).

In the past, there have been various types of DL-based approaches proposed for BI. One is to directly train a deterministic inverse mapping from $\bm{y}$ to $\bm{x}$ \cite{lucas2018using,mccann2017review}. Recent works \cite{ardizzone2019analyzing,andrle2021invertible} proposed to infer the posterior distribution with invertible neural networks (INNs) \cite{kobyzev2020normalizing}. However, INNs require special coupling layers to achieve the normalizing flow, which can be insufficiently expressive and computationally expensive \cite{dinh2014nice}. In addition, a CVAE can be used as as baseline for INN \cite{ardizzone2019analyzing}. This approach, which is denoted by CVAE-vanilla is an oversimplication since $\bm{y}$ and $\bm{z}$ are assumed to be independent, and therefore cannot guarantee accurate estimation of the posterior distribution.

% In addition, the CVAE has been utilized as a baseline method of INN \cite{ardizzone2019analyzing}, which is denoted by CVAE-vanilla. In this paper, we show that CVAE-vanilla \cite{ardizzone2019analyzing}, which is oversimplified by assuming that $\bm{y}$ and $\bm{z}$ are independent, cannot guarantee the performance. 

In order to validate our DL-based approaches for estimating $p(\bm{x}|\bm{y}^*)$, a ground truth is necessary, but it is not available if the only available data is the training dataset, $\{\bm{x}_i,\bm{y}_i\}_{i=1}^D$, and $\bm{y}^*$. We, therefore, used a simple simulation study for dynamic brain PET imaging, in which we can not only generate a training dataset for our DL-based approaches, but also perform MCMC to produce asymptotically unbiased posterior distributions to be used as the gold standard (See Sec. \ref{sec:method} for details). {This simulation study is based on [$^{18}$F] MK-6240, a second generation tau PET tracer.\cite{guehl2019evaluation}} In the simulation, kinetic parameters were first randomly sampled from predefined priors and were then used to generate noisy time-activity curves (TAC) in a target region based on a simplified reference tissue model (SRTM \cite{lammertsma1996simplified}) and Gaussian noise model. For a given testing TAC, the posterior distributions of kinetic parameters using our generative DL-based approaches were compared to the unbiased distributions sampled by MCMC. 
%{Though DL-based approaches take an additional training stage compared to MCMC, they are able to infer posterior distributions $\sim$40 times faster than MCMC in our single-voxel simulation study. Notably, MCMC relies on a defined forward model and is not applicable if only training data are given. In addition, MCMC can be intractable for high-dimensional data.}

\section{Methodology} \label{sec:method}

In this section, we first propose our DL-based methods for posterior estimation. We then describe how we performed the simulation for dynamic brain PET using SRTM and Gaussian noise model. Afterward, we explain in detail how we performed MCMC and our DL-based approaches to obtain the posterior distributions of the kinetic parameters for a given dynamic PET measurement, i.e., a TAC. Finally, we describe how we evaluated the performance of our DL-based approaches using the unbiased posterior distributions sampled by MCMC as the gold standards.

\subsection{DL-based approaches}

In this work, we propose to use a CVAE framework for efficiently sampling the posterior distributions given an observed measurement. We propose different deep neural networks (DNN) for estimating posterior distributions based on the evidence lower bounds (ELBOs) \cite{kingma2013auto,pesteie2019adaptive}.

To estimate the posterior distribution $p(\bm{x}|\bm{y})$ for a given observable measurement $\bm{y}$, we define a random multidimensional latent variable $\bm{z}$ to capture the information loss in the forward process from $\bm{x}$ to $\bm{y}$. We intend to train a neural network (known as decoder), $\theta$, which  performs $\tilde{\bm{x}}=f_{\theta}(\bm{z},\bm{y})$), using dataset $\{\bm{x}_i,\bm{y}_i\}_{i=1}^D$, so that $\tilde{\bm{x}}\sim p(\bm{x}|\bm{y})$ if $\bm{z}$ is sampled from distribution $p(\bm{z}|\bm{x},\bm{y})$, i.e., $\bm{z}\sim p(\bm{z}|\bm{x},\bm{y})$. To make such training possible, we introduce another neural network (known as encoder), $\phi$, which maps $\bm{x}$ and $\bm{y}$ to $\bm{z}$. The two neural networks, i.e., $\theta$ and $\phi$, must be decoupleable after the training, which can be achieved by minimizing the following Kullback-Leibler (KL) divergence:
\begin{align}
  {{KL}(p_{\phi}(\bm{z}|\bm{x},\bm{y})||p(\bm{z}|\bm{x},\bm{y}))}=\int p_{\phi}(\bm{z}|\bm{y},\bm{x})\log \frac{p_{\phi}(\bm{z}|\bm{x},\bm{y})}{p(\bm{z}|\bm{x},\bm{y})} d\bm{z}\nonumber\\
 =\log p(\bm{x}|\bm{y}) - \varepsilon_A = \log p(\bm{x}|\bm{y}) + \log p(\bm{y}) - \varepsilon_B,\label{eq1}
\end{align}
where $\varepsilon_A$ and $\varepsilon_B$ are two equivalent ELBOs, since $\log p(\bm{y})$ is independent to the value of $\bm{z}$. Specifically, we have:
\begin{align}
\varepsilon_A=&\mathbb{E}_{\bm{z}\sim p_{\phi}(\bm{z}|\bm{x},\bm{y})}[\log p_{\theta}(\bm{x}|\bm{y},\bm{z})] - {{KL}(p_{\phi}(\bm{z}|\bm{x},\bm{y})||p_{\phi'}(\bm{z}|\bm{y}))}.\nonumber\\ 
\varepsilon_B=&\mathbb{E}_{\bm{z}\sim p_{\phi}(\bm{z}|\bm{x},\bm{y})}[\log p_{\theta}(\bm{x}|\bm{y},\bm{z})] - {{KL}(p_{\phi}(\bm{z}|\bm{x},\bm{y})||p(\bm{z}))}
+\mathbb{E}_{\bm{z}\sim p_{\phi}(\bm{z}|\bm{x},\bm{y})}[\log p_{\theta'}(\bm{y}|\bm{z})].\label{eq:2}  
\end{align} In the above equation, we replaced $p(\bm{x}|\bm{y},\bm{z})$, $p(\bm{z}|\bm{y})$, and $p(\bm{y}|\bm{z})$ with $p_{\theta}(\bm{x}|\bm{y},\bm{z})$, $p_{\phi'}(\bm{z}|\bm{y})$, and $p_{\theta'}(\bm{y}|\bm{z})$, respectively ($\phi'$ and $\theta'$ represent another encoder and decoder, respectively). In addition, if we assume that $\bm{z}$ is independent of $\bm{y}$, i.e., $p(\bm{z}|\bm{y})=p(\bm{z})$, we have: $\varepsilon_A \approx \varepsilon_C = \mathbb{E}_{\bm{z}\sim p_{\phi}(\bm{x}|\bm{y},\bm{z})} [\log p_{\theta}(\bm{x}|\bm{y,z})] - {{KL}(p_{\phi}(\bm{z}|\bm{x})||p(\bm{z}))}$ and $\varepsilon_B \approx \varepsilon_C + \log p(\bm{y})$. 
 
We therefore propose three different DNNs: CVAE-dual-encoder, CVAE-dual-decoder, and CVAE-vanilla (See Fig. 1), which are designed to maximize $\varepsilon_A$, $\varepsilon_B$, and $\varepsilon_C$, respectively.

\vspace{+5pt}
\subsubsection{CVAE-dual-encoder}
\vspace{+5pt}

Fig. 1 (A) shows the DNN used to maximize $\varepsilon_A$, which consists of an encoder $\phi$ ($[\bm{x},\bm{y}]\rightarrow\bm{z}$), a decoder $\theta$ ($[\bm{y},\bm{z}]\rightarrow \tilde{\bm{x}}$)), and an additional encoder $\phi' (\bm{y}\rightarrow \tilde{\bm{z}})$). 

Maximizing $\mathbb{E}_{\bm{z}\sim p_{\phi}(\bm{z}|\bm{x},\bm{y})}[\log p_{\theta}(\bm{x}|\bm{y},\bm{z})]$ is equivalent to minimizing the following loss function for a training pair:
{\begin{equation}
\begin{aligned}
    \mathcal{L}_{A1}=\frac{1}{2} ||\bm{x} -\tilde{\bm{x}} ||^2_2,\label{eq:3}
\end{aligned}\end{equation}}To maximize $-{{KL}(p_{\phi}(\bm{z}|\bm{x},\bm{y})||p_{\phi'}(\bm{z}|\bm{y}))}$ in $\varepsilon_A$, we use the reparameterization trick \cite{kingma2013auto} in both the encoder neural networks, i.e., $\phi$ and $\phi'$, with two $K$-dimensional multivariable normal distributions defined by $\mathcal{N}({\bm{\mu}}, \text{diag}({\bm{\sigma}^2}))$ and $\mathcal{N}({\bm{\mu}}', \text{diag}({\bm{\sigma}'^2}))$ representing $p_{\phi}(\bm{z}|\bm{x},\bm{y})$ and $p_{\phi'}(\bm{z}|\bm{y})$, respectively. As a result, we introduce another loss function given a pair of training sample:
\begin{align} 
   \mathcal{L}_{A2}={{KL}(p_{\phi}(\bm{z}|\bm{y},\bm{x})||p_{\phi'}(\bm{z}|\bm{y}))}=-\frac{1}{2} \sum^{K}_{k=1}[1+{\rm log}\frac{\bm{\sigma}_{k}^2}{\bm{\sigma}_{k}^{'2}}    - \frac{\bm{\sigma}_{k}^2}{\bm{\sigma}_{k}^{'2}} - \frac{(\bm{\mu}_{k} - \bm{\mu}'_{k})^{2}}{\bm{\sigma}_{k}^{'2}}], \label{eq:4}  
\end{align}where $K$ is the dimension of the latent code $\bm{z}$ or $\tilde{\bm{z}}$. $\bm{\mu}_{k}$ and $\bm{\sigma}_{k}^2$ are the mean and variance of the $k$-th node. In practice, the output layer of each of the two encoders has two branches (each branch consists of $K$ nodes), which represent mean and variance, respectively. We then sample $\bm{z}$ and $\tilde{\bm{z}}$ using $\bm{z}=\bm{\mu}+\bm{\sigma}\odot{\bm{\epsilon}}$ and $\tilde{\bm{z}}=\bm{\mu}'+\bm{\sigma}'\odot{\bm{\epsilon}}$, respectively, where $\bm{\epsilon}$ is a standard multivariable normal distribution, i.e., $\bm{\epsilon}\sim\mathcal{N}(0,\bm{I})$. We then we define the overall loss function as:
\begin{equation}
\begin{aligned} 
   \mathcal{L}_{A}=\mathcal{L}_{A1} + \beta_A\mathcal{L}_{A2} ,
\end{aligned}\label{eq:5}  \end{equation}
where $\beta_A$ is a hyperparameter to weight $\mathcal{L}_{A2}$, defined in Eq. (4).

For inference, given the observation $\bm{y}^*$, we use $\phi'$ to predict $\bm{\sigma}'$ and $\bm{\mu}'$, followed by sampling $\tilde{\bm{z}}$ using $\tilde{\bm{z}}=\bm{\mu}'+\bm{\sigma}'\odot{\bm{\epsilon}}$. Afterward, we concatenate each $\tilde{\bm{z}}$ and $\bm{y}^*$ to the decoder $\theta$ to generate the corresponding $\tilde{\bm{x}}$.

\begin{figure*}[!t]
\centering
\includegraphics[width=16cm]{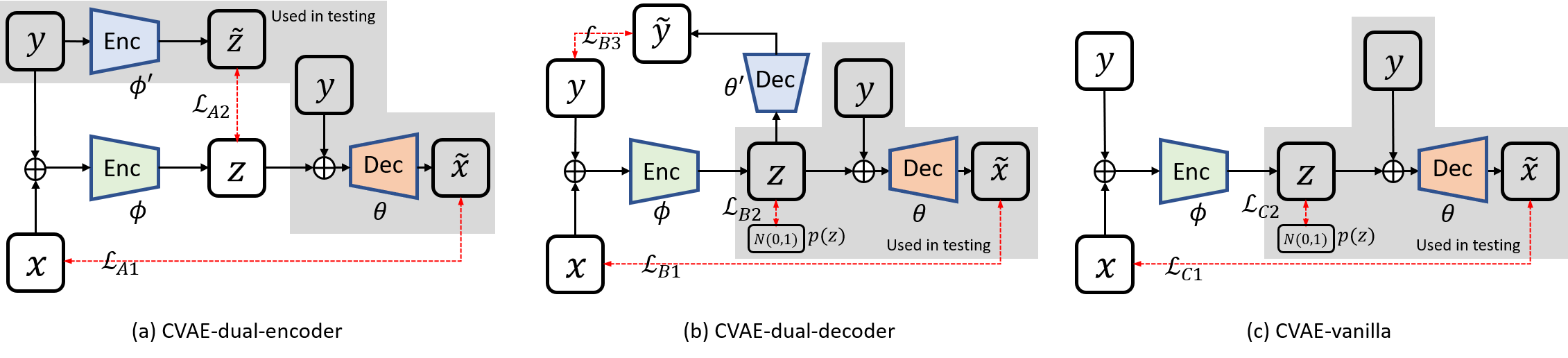}
\caption{Detailed framework of (a) CVAE-dual-encoder, (b) CVAE-dual-decoder and (c) CVAE-vanilla for estimating posterior. Only gray masked decoder is used for inference.}\vspace{+15pt} \label{fig:1}
\end{figure*}

\vspace{+5pt}
\subsubsection{CVAE-dual-decoder}
\vspace{+5pt}

Fig. 1 (B) shows the DNN used to maximize $\varepsilon_B$, which consists of an encoder $\phi$ ($[\bm{x},\bm{y}]\rightarrow\bm{z}$), a decoder $\theta$ ($[\bm{y},\bm{z}]\rightarrow \tilde{\bm{x}}$)), and an additional decoder $\theta' (\bm{z}\rightarrow \tilde{\bm{y}})$).  

Obviously, the first loss function $\mathcal{L}_{B1}$, which is used to maximize the first term in $\varepsilon_B$, is the same as $\mathcal{L}_{A1}$. Similar to CVAE-dual-encoder, we use the same reparameterization trick to handle the KL term in  $\mathcal{L}_{B1}$ with the only difference being the use of $\mathcal{N}(0,\bm{I})$ to represent $p(\bm{z})$. The second loss function becomes
{\begin{equation}
\begin{aligned} 
   \mathcal{L}_{B2}=&  {{KL}(p_{\phi}(\bm{z}|\bm{x},\bm{y})||p(\bm{z}))}=
   - \frac{1}{2} \sum^{K}_{k=1}(1 +{\rm log}(\bm{\sigma}_{k}^2) - \bm{\sigma}_{k}^2 -\bm{\mu}_{k}^2). \label{eq:6}  
  % \frac{1}{2} \sum^{K}_{k=1}(\bm{\sigma}_{k}^2 +\bm{\mu}_{k}^2 -{\rm log}(\bm{\sigma}_{k}^2) -1).
\end{aligned}\end{equation}}We use the following loss function to maximize $\mathbb{E}_{\bm{z}\sim p_{\phi}(\bm{z}|\bm{x},\bm{y})}[\log p_{\theta'}(\bm{y}|\bm{z})]$ in $\varepsilon_B$:
\begin{equation}
\begin{aligned} 
   \mathcal{L}_{B3}=  \frac{1}{2} ||\bm{y}-\tilde{\bm{y}}||^2_2. \label{eq:7} 
\end{aligned}\end{equation} The overall loss function is defined as:
\begin{equation}
\begin{aligned} 
   \mathcal{L}_{B}=\mathcal{L}_{B1} + \beta_B\mathcal{L}_{B2} + \lambda\mathcal{L}_{B3},  \label{eq:8} 
\end{aligned}\end{equation}
{where $\beta_B$ and $\lambda$ are the hyperparameters to weight $\mathcal{L}_{B2}$ and $\mathcal{L}_{B3}$, respectively. Notably, $\mathcal{L}_{B2}$ is different from $\mathcal{L}_{A2}$.}

For inference, we use $\bm{z}\sim\mathcal{N}(0,I)$, and concatenate $\bm{z}$ and $\bm{y}$ to the decoder to generate the corresponding $\tilde{\bm x}$.
 
\vspace{+5pt}
\subsubsection{CVAE-vanilla}
\vspace{+5pt}

Fig. 1 (C) shows the DNN used to maximize $\varepsilon_C$, which consists of an encoder $\phi$ ($[\bm{x},\bm{y}]\rightarrow\bm{z}$) and a decoder $\theta$ ($[\bm{y},\bm{z}]\rightarrow \tilde{\bm{x}}$).  Obviously, we can use $\mathcal{L}_{C1}$, which is the same as both $\mathcal{L}_{A1}$ and $\mathcal{L}_{B1}$, and $\mathcal{L}_{C2}$, which is the same as $\mathcal{L}_{B2}$.
An overall loss function is defined as:
\begin{equation}
\begin{aligned} 
   \mathcal{L}_{C}=\mathcal{L}_{C1} + \beta_C\mathcal{L}_{C2}. \label{eq:9} 
\end{aligned}\end{equation}
where $\beta_C$ is the hyperparameter to weight $\mathcal{L}_{C2}$. For inference, we use $\bm{z}\sim\mathcal{N}(0,\bm{I})$ and concatenate $\bm{z}$ and $\bm{y}$ to the decoder to generate the corresponding $\tilde{\bm{x}}$.

%\in\mathbb{R}^3$ (3 is the number of kinetic parameters)^N$ ($N$ is the number of time frames)

\subsection{Simulation of dynamic PET } 

In dynamic PET, we are interested in estimating posterior distributions of kinetic parameters $\bm{x}$, when a measurement of TAC, $\bm{y}$, is given in a target region. In this study, we used SRTM for tracer kinetics in a brain region, which can be formulated as: 
\begin{eqnarray}
\frac{d C_T(t)}{dt} = R_1 \frac{d C_R(t)}{dt} + k_2C_R(t) - \frac{k_2}{DVR}C_T(t),\label{eq1}
\end{eqnarray}
where $C_T(t)$ and $C_R(t)$ are the activity concentrations in the target region and a pre-defined reference region respectively at time $t$, $DVR$ is the distribution volume ratio between the target and reference region, $k_2$ is the rate constant from free to plasma compartment, and $R_1$ is the ratio of rate constants for transform from plasma to free compartment. The analytic solution of TAC in the target region is:
\begin{eqnarray}
C_T(t) = R_1 C_R(t) + (k_2 - \frac{R_1k_2}{DVR}) C_R(t) \otimes e^{-\frac{k_2}{DVR} t},
\end{eqnarray} 
where $\otimes$ is a convolution operator. As a result, the forward process from kinetic parameters, $\bm{x}=\{DVR, k_2, R_1\}$, to $y_n=\int_{t_{n-1}}^{t_n} C_T(t) \mathrm{d}t + \bm{\epsilon}_n$, where the noise was modeled using $\frac{\bm{\epsilon}_n}{\bm{\sigma}\sqrt{\Delta t_n/T}}\sim \mathcal{N}(0,I)$, $T=\sum_{n=1}^N \Delta t_n$, $\Delta t_n={t_n}-{t_{n-1}}$, and $\bm{\sigma}$ is the standard deviation. We assumed that $\bm{\sigma}$ follows a gamma distribution, i.e., $\bm{\sigma}\sim10^{-4}$Gamma(1,1). For this simulation study, we chose the temporal lobe and cerebellum grey as the target and reference region, respectively. We also set the number of time frames to $N = 54$.
We used the following sequence of time frame durations: 6$\times$10s, 8$\times$15s, 6$\times$30s, 8$\times$60s, 8$\times$120s, and 18$\times$300s. In the simulation, kinetic parameters were randomly sampled from predefined priors (see Sec. II.D) first and then were used to generate noisy TACs in a target region based on a simplified reference tissue model (SRTM) and Gaussian noise model. This noise model is an approximation we made to simplify the simulation because the real noise in PET TACs is actually difficult to characterize. Fig. 2 shows the TAC in the reference region, the TAC without noise in the target region generated using SRTM with $DVR = 1.0$, $k_2 = 0.0006$ min$^{-1}$, and $R_1 = 0.74$, and the TAC with noise by adding Gaussian noise as described.

\begin{figure}[t]
\centering
\includegraphics[width=9cm]{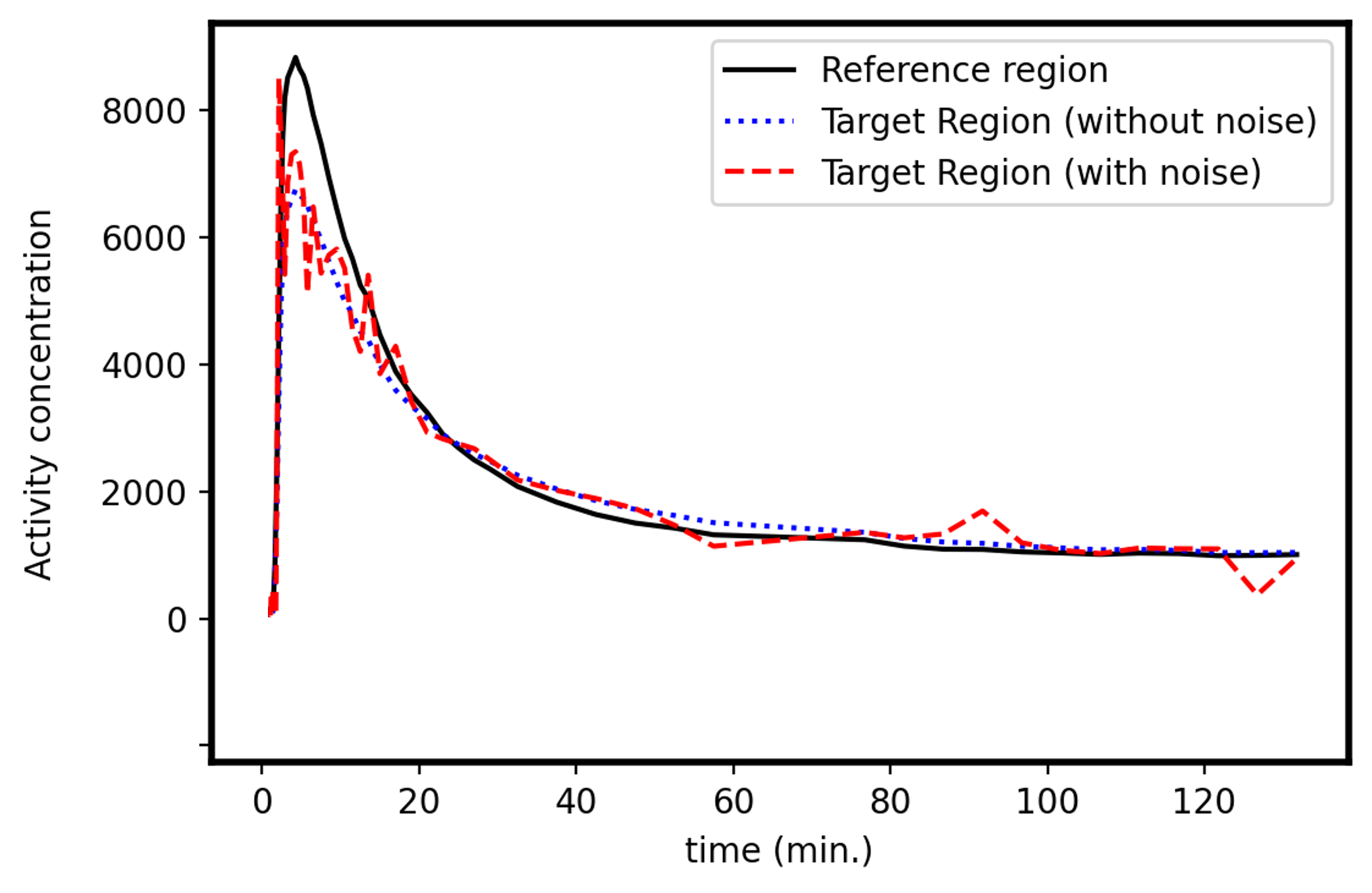}\vspace{-5pt}
\caption{TACs in the reference region and target region (w/wo noise).} \label{fig:2}\vspace{+15pt}
\end{figure}

\subsection{MCMC}

The conventional approach for sampling posterior distribution is to follow a rejection sampling scheme with MCMC \cite{andrieu2003introduction}. If we assume a prior, $p(\bm{x})$, based on our knowledge before the measurement, the posterior distribution is determined as $p(\bm{x}|\bm{y}) \propto  p(\bm{y}|\bm{x})p(\bm{x})$.
In this work, we chose to use the widely used random walk Metropolis-Hastings MCMC (MH-MCMC) to sample the posterior distributions of kinetic parameters. A symmetric proposal distribution that represents a Markov Chain transition from step $l-1$ to step $l$,  $J(\bm{x}{(l)}|\bm{x}^{(l-1)})=N(\bm{x}^{(l)}|\bm{x}^{(l-1)},\sum)$, was used. The diagonal covariance matrix $\sum$ was used to control the acceptance rate of MCMC.

In the implementation of the MCMC, an important step is the judgment of convergence, which usually indicates whether the algorithm is drawing the sample from the true distribution and achieves balance. Trace plots of the (marginal) log-likelihood are often used as visual and subjective tool to give a hint \cite{gelman1996posterior}. To provide a more reliable assessment of convergence, we calculated the mean of the first 10\% and last 50\% steps counted after burn-in steps to check if the difference between these two means is approaching zero \cite{geweke1991evaluating}.

\begin{figure*}[!t]
\centering
\includegraphics[width=17cm]{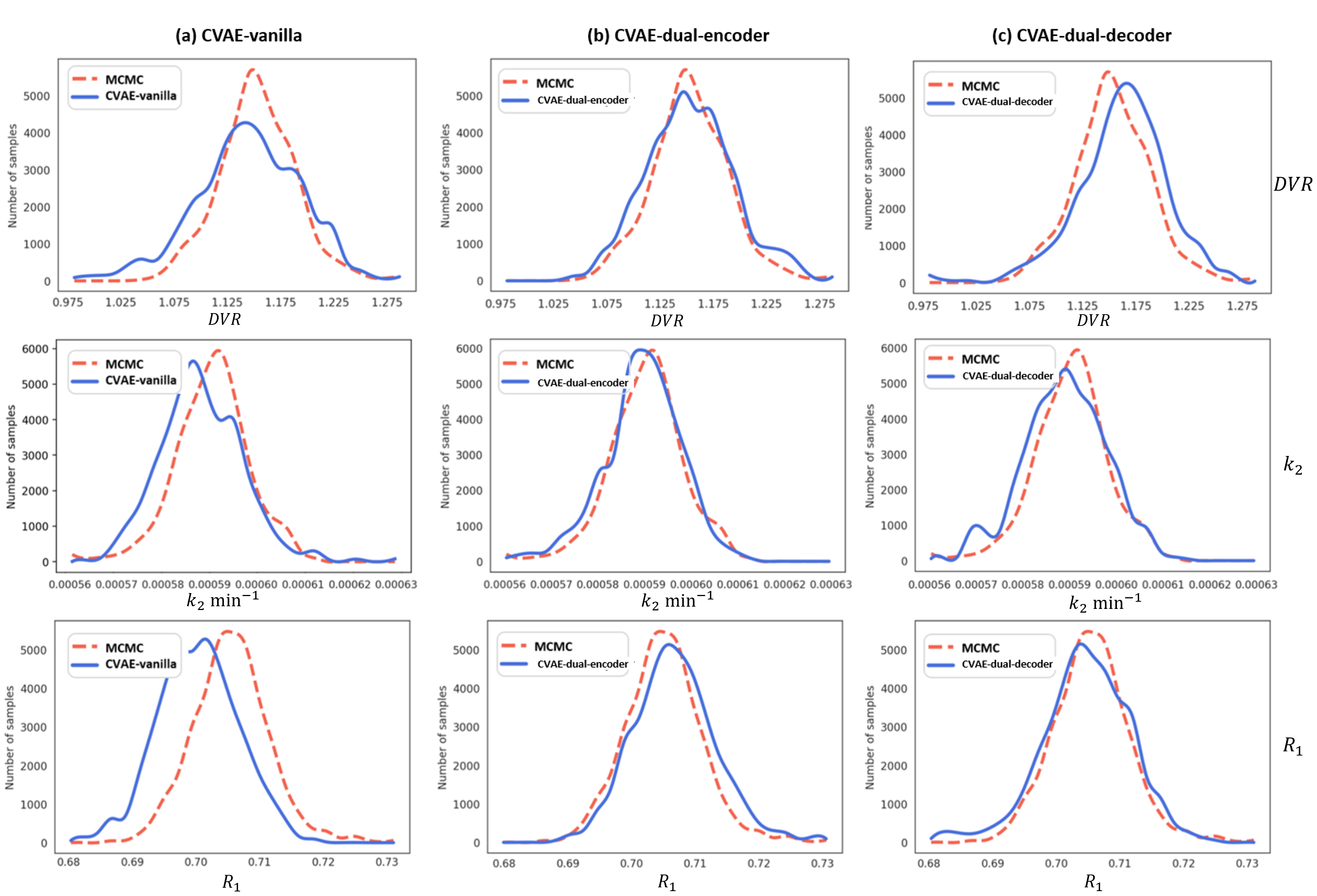} 
\caption{Posterior distributions estimated by MCMC and DL-based approaches for a noisy TAC.} \label{fig:3}\vspace{+15pt}
\end{figure*}

\subsection{Evaluation}

In this study, we first (setting 1) defined the prior $p(\bm{x})$ as  $DVR \sim N(1.0,1.0)$, $k_2\sim N(0.0006\text{min}^{-1},0.01\text{min}^{-1})$, and $R_1 \sim N$(0.74,1.0), based on a previous [$^{18}$F]MK-6240 study across 35 subjects \cite{guehl2019evaluation}. {To demonstrate the effectiveness on multiple simulated kinetic parameter sets, we further increased the mean, variance, and both mean and variance of the prior by 20\%, and denoted as setting {2}, {3}, and {4}, respectively.} For the purpose of quantitative evaluation of our approaches, we kept sampling $\bm{x}= \{DVR,k_2,R_1\}$ from the prior distributions until we collected a total of 200 testing $\bm{x}$’s that satisfy $\frac{|x_i-\tilde{x}_i|}{\tilde{x}_i}< \alpha$,  $i=1,2,3$, where $\tilde{x}_i$ is the mean of prior in each setting, $\alpha=0.26$ was chosen based on the variance of measured $DVR$ across all subjects in the previous study \cite{guehl2019evaluation}. The corresponding testing measurement of TAC, $\bm{y}$, for each testing $\bm{x}$ was then generated using the SRTM and the Gaussian noise model (See II.B).

For each testing measurement, we first used MCMC to sample its corresponding asymptotically unbiased posterior distributions using the defined prior distributions as well as the forward and noise models as described in Sec. II.B. Specifically, in testing, we performed 60,000 iterations of random walk MH-MCMC sampling with 15,000 burn-in steps.
As a result, we generated 45,000 samples for each posterior distribution.

For all our DL-based approaches, we set $\beta=1$ and $\lambda=1$ (only for CVAE-dual-decoder) \cite{kingma2013auto}. We used the same network structure in encoder ${\phi}$ and decoder $\theta$. Specifically, in the encoder, we used four fully connected layers, which contain 128, 100, 50, and 20 nodes. The output layer of the encoder has ten nodes for mean values and ten nodes for variance values, which are in turn used to define the distribution of $\bm{z}$, whose dimension is $K = 10$. In the decoder, we also used four fully connected layers, which contain 128, 100, 50, and 3 nodes. The output layer of the decoder is a three-dimensional vector, i.e., ($DVR$, $k_2$, and $R_1$). In both the encoder and decoder, ReLU is used as an activation function. For the CVAE-dual-encoder, we have an additional encoder ${\phi}'$, which has the same structure as encoder ${\phi}$, though parameters in these two encoders are not shared. For the CVAE-dual-decoder, we had an additional decoder $\theta'$, which has four fully connected layers containing 16, 16, 32, and 54 nodes.

To construct the training set for our DL-based approaches, we generated 10,000 samples of $\bm{x}$ using the defined priors. Each sampled $\bm{x}$ was 
then used to generate its corresponding $\bm{y}$ using the SRTM and the Gaussian noise model. The resulting training data pairs were used to train the neural network in each one of our DL-based approaches. We used the same learning rate, i.e., $10^{-4}$, and stochastic gradient descent (SGD) optimizer with a momentum of 0.9 for all the neural networks. For each approach, the trained neural network was used to generate 45,000 samples to obtain posterior distributions for each testing $\bm{y}$ afterward.

\begin{table}[t]
\centering
\caption{Averaged relative difference of normalized mean  $\overline{\delta}_{\bm{\mu}}$ (\%).}\vspace{+5pt}
\resizebox{0.7\linewidth}{!}{
\begin{tabular}{c|l|c|c|c}
\hline

\multicolumn{2}{l|}{} &CVAE-vanilla& CVAE-dual-encoder& CVAE-dual-decoder\\\hline\hline

&$DVR$  &10.5 & 8.3 & 8.3 \\
Set {1}&$k_2$  (min$^{-1}$) &13.8 & 11.9 &  11.6 \\
&$R_1$  &8.5  & 7.1   & 7.2 \\\hline\hline

&$DVR$  &10.2 & 8.5 & 8.4 \\
Set {2}&$k_2$  (min$^{-1}$) &13.2 & 11.1 &  11.3 \\
&$R_1$  &8.5  & 7.1   & 7.0 \\\hline\hline

&$DVR$  &11.2 & 8.7 & 8.9 \\
Set {3}&$k_2$  (min$^{-1}$) &13.9 & 12.3 &  12.0 \\
&$R_1$  &8.9  & 7.7   & 7.4 \\\hline\hline

&$DVR$  &10.7 & 8.4 & 8.6 \\
Set {4}&$k_2$  (min$^{-1}$) &13.4 & 11.7 &  11.8 \\
&$R_1$  &8.7  & 7.5   & 7.6 \\ 
\hline
\end{tabular}}
\label{tabel:1}\vspace{+15pt} 
\end{table}

To evaluate the performance of each DL-based approach using MCMC as the reference, we first computed the average relative difference of normalized mean and standard deviation, i.e., $\overline{\delta}_{\bm{\mu}}$ and $\overline{\delta}_{\bm{\sigma}}$, across $M=200$ testing samples for each kinetic parameter using:
\begin{equation}
\begin{aligned} 
   \overline{\delta}_{\bm{\mu}} &= \frac{1}{M}\sum_m\frac{|\bm{\mu}_m^{MCMC}-\bm{\mu}_m^{DL}|}{\bm{\mu}_m^{MCMC}},\\
   \overline{\delta}_{\bm{\sigma}} &= \frac{1}{M}\sum_m\frac{|\bm{\sigma}^{MCMC}_m-\bm{\sigma}_m^{DL}|}{\bm{\sigma}_m^{MCMC}},\label{eq:12}
\end{aligned}\end{equation}
where $\{\bm{\mu}_m^{MCMC},\bm{\sigma}_m^{MCMC}\}$ ($\{\bm{\mu}_m^{DL},\bm{\sigma}_m^{DL}\}$) are the mean and standard deviation obtained by fitting the corresponding posterior distribution from MCMC (DL-based approach) for the $m$-th sample using a Gaussian function. We also computed the average KL divergence, $\overline{D}$, across all the testing samples using:
\begin{equation}
\begin{aligned} 
   \overline{D} &= \frac{1}{M} \sum_m D_{KL}(p_m^{MCMC}(\bm{x}|\bm{y})||p_m^{DL}(\bm{x}|\bm{y})),\label{eq:13}
\end{aligned}\end{equation}where $p_m^{MCMC}$ and $p_m^{DL}$ are the posterior distributions from MCMC and the DL-based approach, respectively, for the $m$-th observable testing sample.

We used the PyTorch toolbox for the implementation of our DL-based approaches. {We performed all the computation on a server with an NVIDIA V100 GPU (32GB graphics RAM version) and an Intel Xeon 8-core CPU alongside 24GB of RAM.}

\begin{table}[t]
\centering
\caption{Averaged relative difference of normalized standard deviation $\overline{\delta}_{\bm{\sigma}}$ (\%).}\vspace{+5pt}
\resizebox{0.7\linewidth}{!}{
\begin{tabular}{c|l|c|c|c}
\hline

\multicolumn{2}{l|}{}&CVAE-vanilla& CVAE-dual-encoder& CVAE-dual-decoder\\\hline\hline

&$DVR$  &9.4 & 7.1 & 6.6 \\
Set {1}&$k_2$  (min$^{-1}$)  &8.4 & 6.0 &  6.3 \\
&$R_1$  &12.7  & 10.4   & 10.2 \\\hline\hline

&$DVR$  &9.8 & 7.4 & 7.2 \\
Set {2}&$k_2$  (min$^{-1}$)  &8.6 & 6.7 &  6.9 \\
&$R_1$  &12.8  & 10.6   & 10.3 \\\hline\hline

&$DVR$  &9.1 & 6.8 & 6.5 \\
Set {3}&$k_2$  (min$^{-1}$)  &8.1 & 6.2 &  6.0 \\
&$R_1$  &12.6  & 10.6   & 10.4 \\\hline\hline

&$DVR$  &9.5 & 7.6 & 7.1 \\
Set {4}&$k_2$  (min$^{-1}$)  &8.7 & 6.4 &  6.4 \\
&$R_1$  &12.8  & 10.6   & 10.6 \\ 

\hline 
\end{tabular}}
\label{tabel:2}\vspace{+15pt} 
\end{table}

\begin{table}[t]
\centering
\caption{Averaged KL divergences between the posterior inferenced by MCMC and the proposed CVAEs.}\vspace{+5pt}
\resizebox{0.7\linewidth}{!}{
\begin{tabular}{c|l|c|c|c}
\hline

\multicolumn{2}{l|}{}&CVAE-vanilla& CVAE-dual-encoder& CVAE-dual-decoder\\\hline\hline

&$DVR$  &0.107 & 0.078 & 0.075 \\
Set {1}&$k_2$  (min$^{-1}$) &0.143 & 0.103 & 0.093 \\
&$R_1$  &0.125 & 0.062 & 0.085 \\\hline\hline

&$DVR$  &0.110 & 0.091 & 0.082 \\
Set {2}&$k_2$  (min$^{-1}$) &0.146 & 0.105 & 0.102 \\
&$R_1$  &0.127 & 0.069 & 0.083 \\\hline\hline

&$DVR$  &0.109 & 0.088 & 0.086 \\
Set {3}&$k_2$  (min$^{-1}$) &0.141 & 0.108 & 0.104 \\
&$R_1$  &0.125 & 0.062 & 0.085 \\\hline\hline

&$DVR$  &0.102 & 0.080 & 0.079 \\
Set {4}&$k_2$  (min$^{-1}$) &0.142 & 0.098 & 0.094 \\
&$R_1$  &0.120 & 0.075 & 0.082 \\

\hline 
\end{tabular}}
\label{tabel:3}\vspace{+15pt} 
\end{table}

\vspace{+15pt}  
\section{Results and Discussions}

Fig. 3 shows posterior distributions of $DVR$, $k_2$, and $R_1$ obtained from MCMC and DL-based methods for a single noisy TAC measurement $\bm{y^*}$ generated using $DVR = 1.0$, $k_2 = 0.0006$ min$^{-1}$, and $R_1 = 0.74$. All DL-based approaches agree reasonably well with asymptotically unbiased MCMC, while both CVAE-dual-encoder and CVAE-dual-decoder yield better agreement than the CVAE-vanilla.

Tables 1-3 show $\overline{\delta}_{\bm{\mu}}$, $\overline{\delta}_{\bm{\sigma}}$, and $\overline{D}$, respectively, for each kinetic parameter and each DL-based approach. All the results show that both CVAE-dual-encoder and CVAE-dual-decoder yield better agreement with MCMC than CVAE-vanilla. For both the relative shifts of mean and standard deviation of the posterior distribution for each kinetic parameter, CVAE-dual-encoder and CVAE-dual-decoder outperform CVAE-vanilla by $\sim$2\%, which is expected because CVAE-vanilla is an approximation of both CVAE-dual-encoder and CVAE-dual-encoder as described in Sec. II.A.

Fig. 4 shows the average KL divergence $\overline{D}$, across 200 testing samples as defined in Eq. \ref{eq:13} versus hyperparameters $\beta_A$, $\beta_B$ and $\lambda$ in CVAE-dual-encoder and CVAE-dual-decoder. We note that all of the standard deviations for $\overline{D}$ are measured over three network training runs. It appears that $\overline{D}$ is insensitive to all the hyperparameters (i.e., $\beta_A$, $\beta_B$ and $\lambda$) for the same range of [0.6,1.8]. Therefore, we simply use $\beta_A=1$, $\beta_B=1$ and $\lambda=1$ for all of our DL-based approaches.

\begin{figure}[!t]
\centering
\includegraphics[width=17cm]{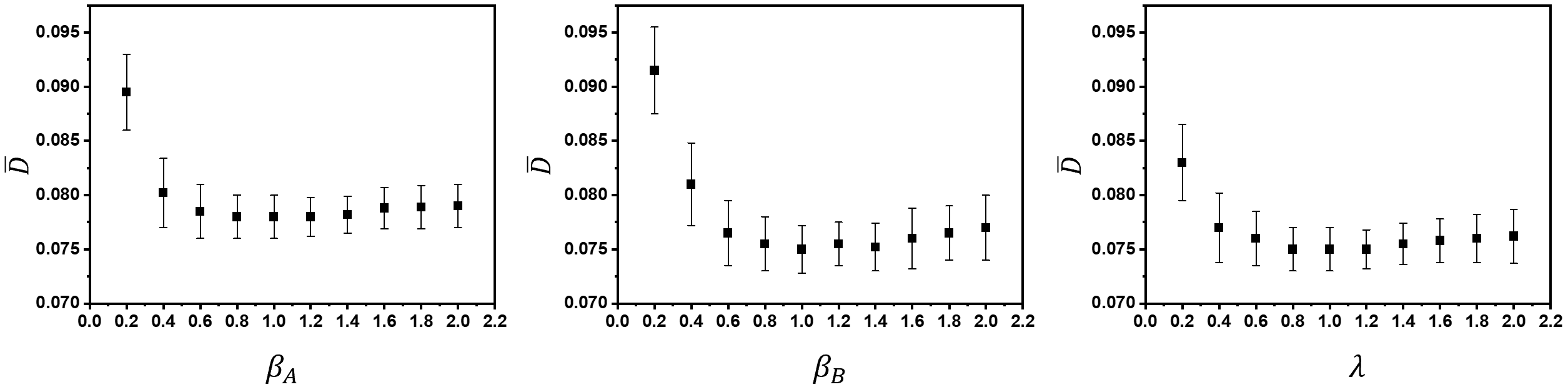} 
\caption{The sensitivity analysis of $\beta_A$ in our CVAE-dual-encoder framework, and $\beta_B$, $\lambda$ in our CVAE-dual-decoder framework.} \vspace{+20pt}\label{fig:4}
\end{figure}

{Fig. 5 shows $\overline{D}$ versus the number of the training samples for CVAE-dual-decoder. $\overline{D}$ reaches a plateau if the number of training samples is more than $\sim$4000. The uncertainty (as shown by the error-bar size for each data point in the figure), which is quantified by the standard deviation of $\overline{D}$ across all 200 testing samples, also decreases and then remains relatively constant as the number of training samples increases. When we consider that the uncertainty is composed of aleatoric and epistemic components, the epistemic uncertainty becomes smaller as we increase the number of training samples and the aleatoric uncertainty dominates when the number of training samples is more than $\sim$5000.}

\begin{figure}[!t]
\centering
\includegraphics[width=8cm]{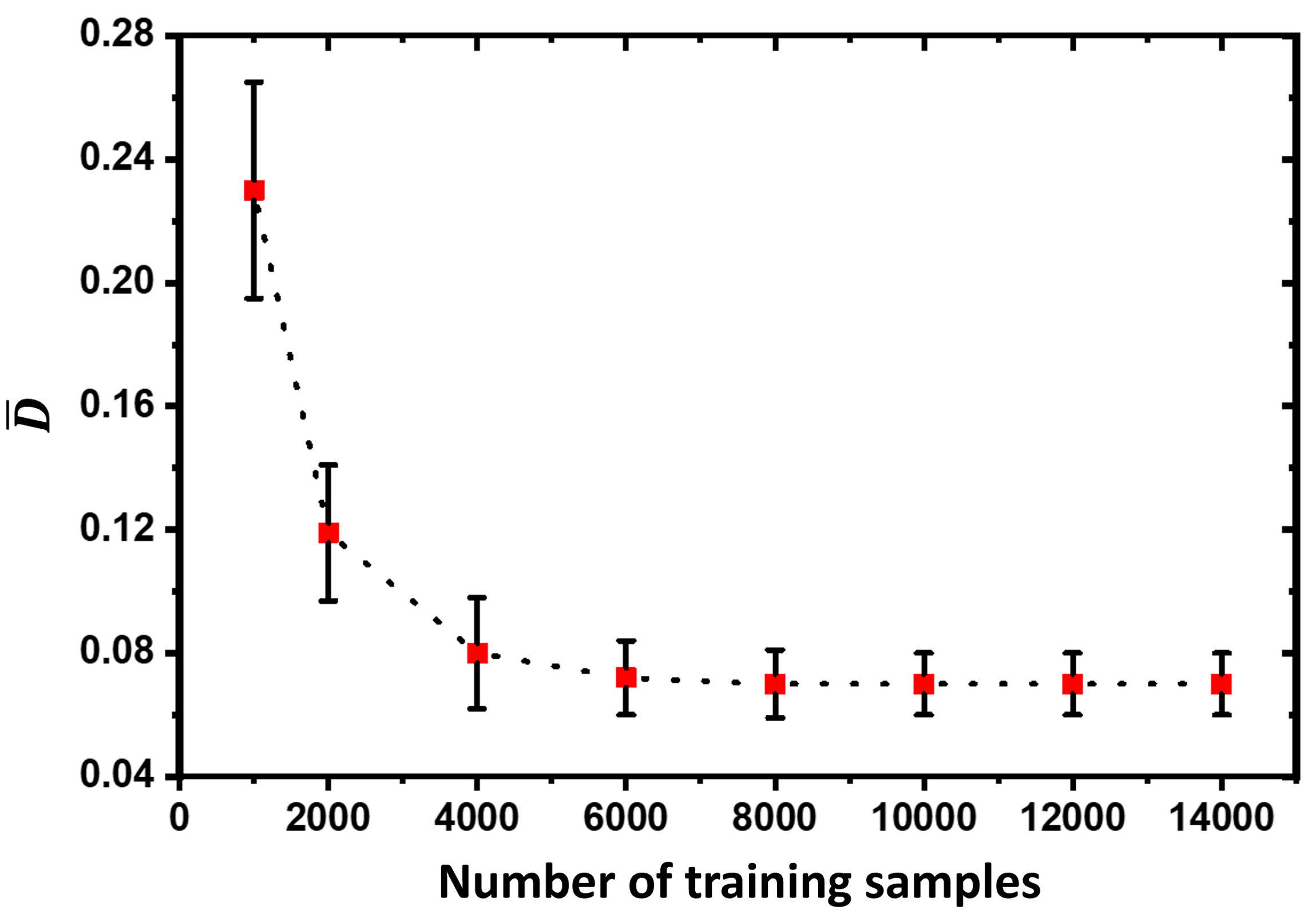} 
\caption{Average KL divergence ($\overline{D}$) versus the number of training samples.}\vspace{+15pt} \label{fig:5}
\end{figure}

It took $\sim$10 minutes to sample 45,000 samples using PyMC~\cite{salvatier2016probabilistic}, a python based MCMC implementation. All the neural networks in our DL-based approaches were trained with 200 epochs, which took $\sim$2, 2, and 2.5 hours for CVAE-vanilla, CVAE-dual-decoder, and CVAE-dual-encoder, respectively. It took less than 15 seconds for each trained CVAE neural network to infer posterior distributions for all three parameters (with 45,000 samples) for a given $\bm{y}^*$. For estimating posterior distributions for a problem where MCMC is feasible, our DL-based approaches can be, therefore, much more efficient than MCMC. In this study, we focused on a single-region dynamic brain PET, which allows us to perform both MCMC and our DL-based approaches for estimating posterior distributions. For a problem with high-dimensional data, MCMC  can be computationally intractable, while a trained neural network is still feasible.

In the following subsections, we provide some detailed discussions on a few topics related to MCMC and DL-based methods for estimating posterior distributions.

\noindent\underline{Convergence of MCMC}

A critical question in MCMC is to determine if the sampling has converged to a stationary distribution. Though several convergence criteria have been proposed in the past, in practice, the convergence is often determined empirically, e.g., using Geweke's test based on the trace of temporal series \cite{geweke1991evaluating}. Fig. 6 shows a temporal series of $DVR$ values sampled by MCMC. After 15,000 burn-in steps,
the sampled values become stable. For all three parameters (i.e., $DVR$, $k_2$, and $R_1$), with 15,000 burn-in steps, the difference between the mean values from the first 10\% and from the last 50\% steps are less than 0.001, which indicates a good convergence \cite{geweke1991evaluating}.

\begin{figure}[!t]
\centering
\includegraphics[width=9cm]{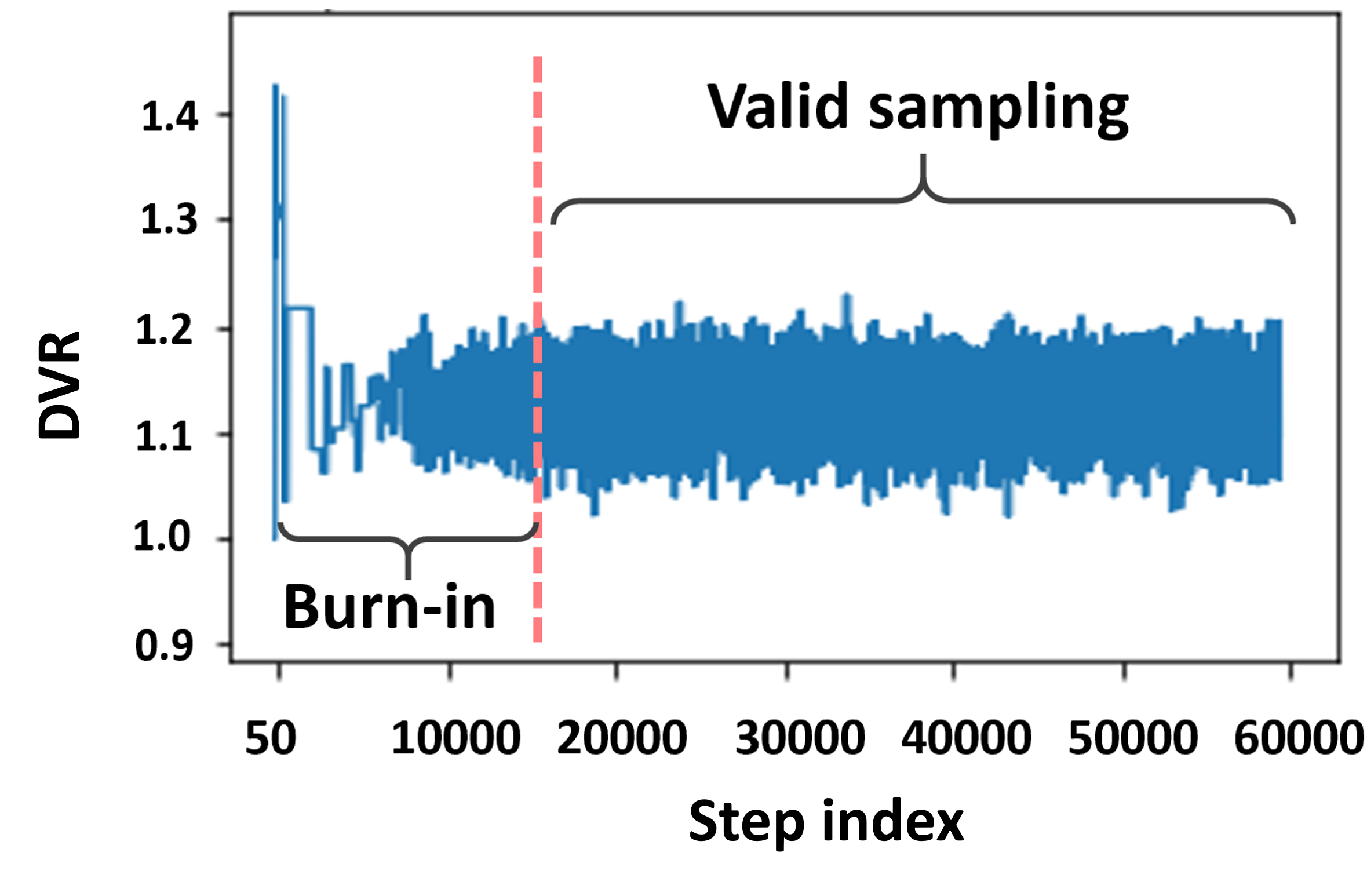} 
\caption{A temporal series of the samples drawn for $DVR$ using MCMC.} \vspace{+15pt} \label{fig:6}
\end{figure}

\noindent\underline{Mismatch between training and testing data}

For all the DL-based methods, data shift represents a mismatch between the distributions of the training and testing data, which can degrade the performance. For this simulation study, we expect that the performance, measured using $\overline{D}$ as defined in Eq. \ref{eq:13}, to deteriorate if the probability of the $DVR$ value used to generate the measurement, $\bm{y}^*$, is low based on the prior distribution of $DVR$ used to generate the training data. 
Fig. 7 shows $\overline{D}$ versus ${DVR}^*$, which is the $DVR$ value used to generate $\bm{y}^*$ measurements. We can see that $\overline{D}$ increases relatively slowly versus ${DVR}^*$ if ${DVR}^*<3.35$, i.e., the sum of mean (i.e., 1) and full width at half maximum (FWHM) of $\mathcal{N}(1,1)$ (i.e., 2.35). The increase of $\overline{D}$ relative to ${DVR}^*$ becomes much faster if ${DVR}^*>3.35$. As a result, when we apply our DL-based methods to the problem as defined in Sec. II, the prior distribution, $p(\bm{x})$, which is implicitly defined by the training data, should cover the value of $\bm{x}$ corresponding to measurement $\bm{y}^*$.

\noindent\underline{CVAE-dual-encoder vs CVAE-dual-decoder} 

As shown in Fig. 3 and Tables 1-3, CVAE-dual-encoder and CVAE-dual-decoder have similar performance.  This is expected because both methods are essentially equivalent from the perspective of variational inference without approximation. They both have three network modules to be trained (CVAE-dual-encoder: 2 encoders, 1 decoder; CVAE-dual-decoder: 1 encoder, 2 decoders).  

Compared to CVAE-dual-encoder, CVAE-dual-decoder requires more training time because it has one more loss term. However, CVAE-dual-decoder is faster in inference because it has a simpler inference structure than CVAE-dual-encoder (See the gray areas in Fig. 1). The user may choose either of them for the specific task based on their characteristics described above.

\begin{figure}[!t]
\centering
\includegraphics[width=8cm]{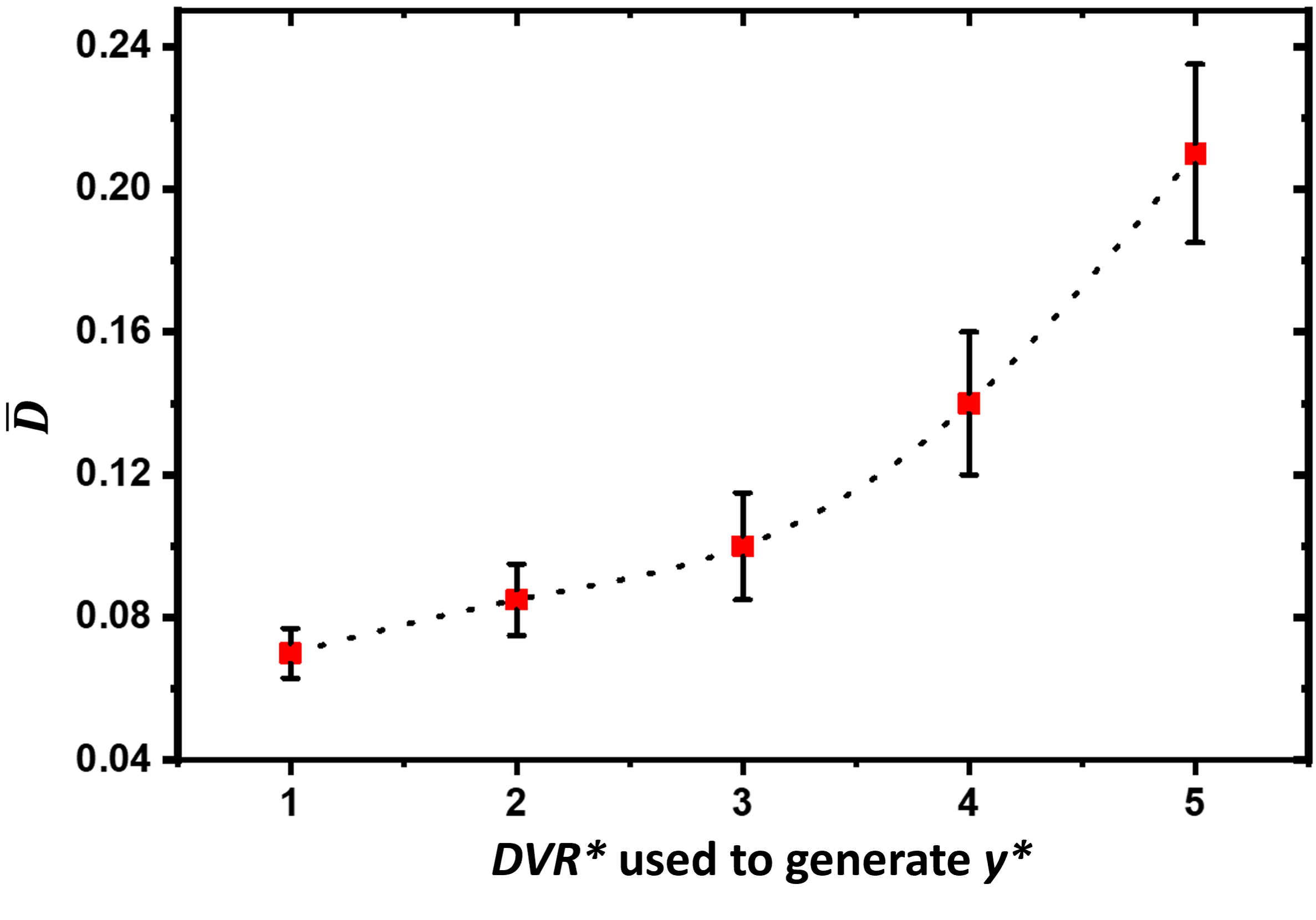} 
\caption{Average KL divergence $\overline{D}$ versus $DVR^*$'s used to generate testing $\bm{y}*$.}\vspace{+15pt}  \label{fig:7}
\end{figure}

\noindent\underline{{Future work}}

{We would like to point out that our goal is to use deep learning to solve the general problem as stated in Sec. I. In this problem, we assume that the training data are already available to us. We applied our deep learning approaches to dynamic brain PET ($\bm{x}$: kinetic parameters, $\bm{y}$: TACs) that can be described by SRTM. In such a problem, we can not only generate the data for the training of deep neural networks but also perform MCMC. As a result, we are able to evaluate the performance of our deep learning approaches using MCMC posterior distributions as reference. For this particular problem, both MCMC and one of our deep learning approaches can be used to estimate posterior distributions for a given measurement (i.e., TAC) on a subject (a definition of prior of kinetic parameters is needed for both approaches).} {It is also worth noting that source and target domains should be the same for our deep-learning based approaches to avoid inference bias\cite{liu2022deep}. For example, unless some domain adaption techniques are used, it is not appropriate to apply a neural network trained for one tracer to obtain posterior distributions for a different tracer because priors, for example, can be very different for different tracers even the same kinetic model is used.}  

%If we can use the same kinetic model, for example, SRTM, and the same priors for two different tracers, a neural network trained with one tracer can be applied to the other.  However, a reasonable prior for a tracer is usually defined based on previous PET studies using the same tracer. For example, DVR prior in a target brain region for tau-PET can be a Gaussian distribution, whose mean and standard deviation are quantified by a previous study that includes many subjects. Different tracers may have very different priors. As a result, even if the same kinetic model is used, we may still need to perform training for each tracer separately.

In this work, we estimated the posterior distributions of kinetic parameters given a measurement of TAC, i.e., $\bm{y}^*$, in dynamic brain PET. We can, for example, extend our work so that $\bm{y}$ represents dynamic sinogram data rather than a TAC. As stated in Sec. I, our DL-based approaches for estimating posterior distributions are general and can be applied to many medical applications.

\section{Conclusions}

We have proposed DL-based approaches for estimating posterior distributions. Our approaches, which are based on a deep variational inference framework, are implemented using two different deep neural networks, CVAE-dual-encoder and CVAE-dual-decoder. The conventional CVAE framework, i.e., CVAE-vanilla, can be regarded as a simplified case of these two neural networks. All these neural networks have different characteristics and can be chosen by the user for specific applications. We have applied these approaches to a simulation study of dynamic brain PET and evaluated their performance using asymptotically unbiased MCMC as the reference. Both CVAE-dual-encoder and CVAE-dual-decoder yield good agreement with MCMC for estimating posterior distributions of kinetic parameters given a measurement of TAC. For our simulation study, we have also found that CVAE-vanilla can also be used for estimating posterior distributions, although it has an inferior performance to both CVAE-dual-encoder and CVAE-dual-decoder.

\section*{Acknowledgments}

This work is supported in part by NIH P41EB022544.

\section*{Conflicts of interest or financial disclosures}

The authors have no conflicts to disclose.

\section*{Data availability statement}

This paper is a simulation study. All of the data are synthesized with the SRTM model detailed in the main text.

\newpage

% following only if there is an appendix

\section*{References}
\addcontentsline{toc}{section}{\numberline{}References}
\vspace*{-20mm}

% Following assumes you are using bibtex. However, for submission to the
% journal you MUST explicitly INCLUDE THE REFERENCES IN THE TEX FILE. 
% In that case you need the following

% \begin{thebibliography}{10}
% insert the .bbl file generated by bibtex here
	%This will be a series of entries from your .bib file formatted
	%something like
	%\bibitem{Me09}
        %{I.~Meijsing, B.~W.~Raaymakers, A.~J.~E.~Raaijmakers \it et al.},
        %\newblock {Dosimetry for the MRI accelerator: the impact of a 
	%magnetic field on the response of a Farmer NE2571 ionization chamber},
        %\newblock Phys. Med. Biol. {\bf 54}, 2993 -- 3002 (2009).

% \end{thebibliography}

% The following is when using bibtex and picks up the example.bib file

%\bibliography{Explicit address of .bib file}
\bibliography{example}      %example.bib is on the same directory
% above points to where we find the master reference list
% and also causes the bibliography to be printed

% When creating your bibliography you should run bibtex on your local
% computer after running pdflatex on your .tex file. bibtex will
% generate a .bbl file.
% Copy the contents of this .bbl file into your main latex document,
% replacing the "\bibliography" command which was pointing at your .bib file.

% following defines style of .bbl file 

%\bibliographystyle{explicit relative path to medphy.bst}
\bibliographystyle{medphy.bst}    %if this is installed on your system,
				    %it is not essential to have the    ./

\end{document}